\documentclass[12pt,titlepage]{article}
\usepackage[pdftex]{graphicx}
\usepackage[numbers]{natbib}

\begin{document}

\title{Galaxy Cluster Astrophysics and Cosmology: Questions and Opportunities for the Coming Decade}

\author{S.T.~Myers (NRAO)\thanks{contact author: 
  P.O.~Box O, Socorro, NM, 87801; email: {\tt smyers@nrao.edu}}, 
  C.~Pfrommer (CITA), \\ 
  J.~Aguirre (Penn), J.R.~Bond (CITA),  J.O.~Burns (Colorado), \\ 
  T.~Clarke (NRL), M.~Devlin (Penn), A.~Evrard (UMich), \\ 
  S.~Golwala (Caltech), S.~Habib (LANL), K.~Heitmann (LANL), \\ 
  W.L.~Holzapfel (UCBerkeley), N.E.~Kassim (NRL), A.~Kravtsov (Chicago), \\
  A.T.~Lee (UCBerkeley), M.~Markevich (CfA), D.~Marrone (Chicago), \\
  D.~Nagai (Yale), L.~Page (Princeton), E.~Pierpaoli (USC)\\, 
  L.~Rudnick (UMinn), J.~Sievers (CITA), G.~Taylor (UNM), M.~Voit (MSU)
}

\date{{\it Astro2010 Science White Paper} (2009-02-15) \\[3ex]
\begin{minipage}[h]{6.5in}
  \begin{center}
  \includegraphics[width=6.5in]{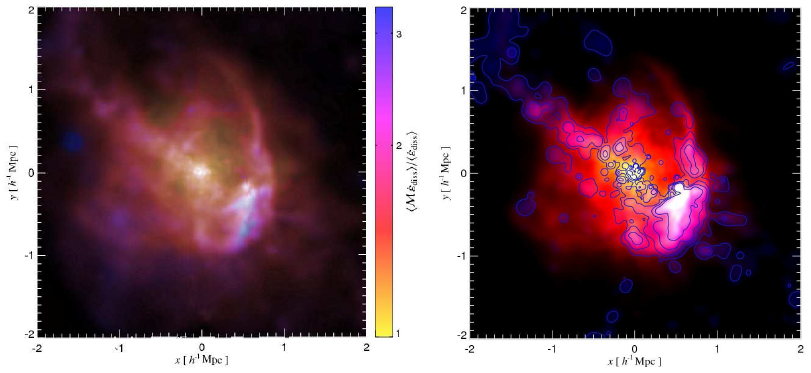} \break
  {\scriptsize \it Simulated
  cluster shocks with Mach numbers (L) and 150MHz emission (R). 
  Fig~1 from Battaglia et al.\ (2008).}
  \end{center}
\end{minipage}
}

\maketitle

\begin{center}
{\Large \sc Galaxy Cluster Astrophysics and Cosmology: \break
 Questions and Opportunities for the Coming Decade}
\end{center}



\section{Motivation and questions}

Clusters of galaxies provide us the opportunity to study an ``ecosystem'' --- a
volume that is a high-density microcosm of the rest of the Universe.  Clusters
are signposts for early structure formation, and are moderately isolated,
growing on the Hubble timescale from the Cosmic Web.  Clusters are excellent
laboratories for studying plasma physical processes as well as for studying how
super-massive black holes interact with the ambient cluster plasma.  The next
generation of cluster surveys is well suited to address fundamental problems in
physics and cosmology (e.g.\ \cite{q2c}), such as further constraining the Dark
Energy equation of state \cite{detf} or to test whether our understanding of
gravity is complete.  Time is ripe to tackle the following important questions
with clusters on an individual basis or as an entire population using
multi-wavelength observational campaigns:

\begin{list}{$\bullet$}{
  \setlength{\topsep}{0ex}\setlength{\itemsep}{0ex plus0.2ex}
  \setlength{\parsep}{0.5ex plus0.2ex minus0.1ex}}

\item \textbf{How do clusters form and grow?}  How does
  feedback from star, galaxy, and black hole formation impact cluster structure and
  evolution?  What detailed physical processes govern the heating and cooling of
  cluster cores?  Are clusters pre-heated before they are assembled?
  What are the robust mass observables and scaling relations that can
  be used for cosmology?

\item \textbf{How do the cluster medium and its constituents evolve?}  How
  much pressure is provided by the thermal plasma, by turbulence, and
  by the non-thermal particle populations?  
  How are metals mixed into the medium?  
  How does the hot ICM affect evolution of cluster galaxies and vice
  versa?  

\item \textbf{How do we use clusters of galaxies as a window on fundamental
  physics?}  Does Dark Matter interact or is it collisionless? Does
  Dark Matter annihilate? 
  How does Dark Energy affect the growth and evolution of clusters? Does Dark
  Energy only change the expansion history of the Universe or is our
  understanding of gravity on the largest scales of the Universe incomplete?
  Do exotica (cosmic strings) impact cluster and structure formation in any
  way?

\item \textbf{What can we learn about plasma astrophysics in clusters?}  What
  causes non-thermal high energy cluster emission in the radio and hard X-rays? What is
  the origin of large scale magnetic fields and how do they evolve?  How do
  shocks of moderate strength accelerate relativistic particles?  What are the
  properties of turbulence in the cluster medium, and how is this coupled to
  the cluster structure? Do anisotropic transport processes in the
  collisionless plasma play an important role?

\end{list}

These questions are closely interlinked and require a multifaceted approach
bridging the observational and theoretical. 
Complementary to other approaches to measuring the expansion history
of the Universe (e.g. SNeIa, BAO), future large surveys of galaxy clusters
potentially provide a powerful probe for the growth of structure and
hence are an invaluable tool for testing our understanding of General Relativity. This
however is only possible if systematics associated with cluster mass
calibrations, sample selection effects, and our incomplete knowledge of the
physics of the intracluster medium can be understood and controlled.


\begin{figure}[t!]
  \includegraphics[width=\textwidth]{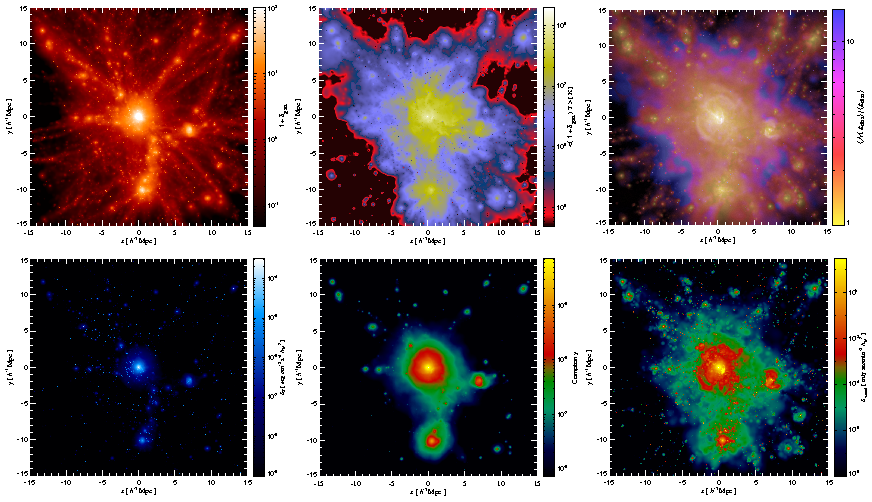}
  \caption{\label{fig:g72} Simulated super-cluster region around a Coma-like
    cluster. {\em Top row L to R}: Projected gas density, mass-weighted
    temperature, and Mach number of formation shocks. 
    {\em Bottom row L to R:} Associated observables 
    X-ray surface brightness, amplitude of the SZ effect, and
    the radio synchrotron emission. Note that each of these observables is
    sensitive to different physical properties and combining different
    observables enables us to solve for the underlying physics.  Taken from
    \protect\cite{Pfrommer08}.  }
\end{figure}

Clusters of galaxies provide unique opportunities for investigating
non-equilibrium processes in plasma astrophysics. Combining 
cluster observables from the radio to Gamma-rays enables
us to probe unique characteristics of the intracluster plasma. 
Conditions range from magnetically dominated regions within radio lobes to
plasma dominated by thermal pressure, with bulk motions 
being subsonic within cooling cores to predominantly supersonic in the
accretion regions.  Hence, observationally and theoretically, we
expect a complex interplay between cosmic rays, magnetic fields, and turbulence
which poses a range of intriguing physical puzzles. There are many excellent
recent papers on observational and theoretical aspects of cluster
astrophysics and cosmology, e.g.\ 
\cite{Voit}.\footnote{In this white paper, unfortunately, we do not have the
  space to do proper justice to all the authors and papers on this subject, and
  only provide a few examples as references.  We have also shown a
  modest bias towards papers and examples from the authors' work, and for
  highlighting the most recent results.}

\section{Cluster Astrophysics Now}

In the past decade, great advances have been made in our understanding of
clusters of galaxies, from the standpoint of their internal structure and
evolution to their place in the larger scale structure of the Universe.
These advances have been made possible by tremendous improvements in
theoretical modeling and numerical simulation, as well as a wealth of
new information provided by multi-wavelength surveys of the Universe.
Below is a list of highlights that are jumping off points for future work:

\noindent\underline{\it Explosion of Cluster Observations:} 
SDSS and other optical/IR galaxy surveys
have yielded many new candidates. The {\sc ROSAT} cluster surveys are still our
best all-sky X-ray sample, and {\sc Chandra} and {\sc XMM-Newton} have
provided spectacular detailed images and spectra.
Progress is still being made in the use of these surveys for
characterizing the growth and evolution of clusters through cosmic time.  
Future O/IR and X-ray surveys will greatly expand these classic
cluster catalogs, while new samples will be obtained from the
identification of clusters through the Sunyaev-Zel'dovich (SZ) effect
using bolometer cameras on millimeter and submillimeter telescopes.  
A combined analysis of X-ray,
Sunyaev-Zel'dovich, and weak-lensing data of a large cluster sample is needed
to address possible biases, e.g.\ \cite{Mahdavi07}.

\noindent\underline{\it New Simulation Technology:} 
There are a number of excellent and widely-used
N-body and hydrodynamic simulation codes that have been developed throughout
the past decade by the theoretical community (e.g. \cite{gadget}).  These,
along with analytic and semi-analytic modeling, have ushered in a new era in
our understanding of the astrophysics of clusters and large scale structure, as
well as enabled us to feed simulated Universes into our designs for the
next-generation of astronomical facilities.  See Figure~\ref{fig:g72}.
Another example is the ``Millennium Simulation''\footnote{
  {\tt http://www.mpa-garching.mpg.de/galform/millennium/}}, 
one of many widely-used resources for astrophysics and cosmology.
Extensive comparisons between codes have been carried out
\cite{cccp} to verify the validity of results.

\noindent\underline{\it Parameter Extraction:} 
As simulations become increasingly realistic, they also become more
complex, and defined by a larger number of physical and computational
parameters. Determining these parameters by confronting a necessarily
finite number of simulations with observational data will be a future
challenge. Techniques for addressing this problem are being developed
\cite{habib07}.

\noindent\underline{\it Evolution:} 
Using cluster samples to study cosmology as well as the
evolution of structure requires understanding the relation between the
observables (X-ray luminosity, Compton-$y$, richness) and the actual masses and
structure of the clusters.  Current work has been focusing on calibration of
scaling relations (e.g.\ \cite{Rykoff08}) but it will be an important task to
push this to higher redshifts.

\noindent\underline{\it Internal Structure and Shocks:} 
Sensitive X-ray observations of clusters using the 
new generation of space telescopes {\sc Chandra} and {\sc XMM-Newton} are
extending our knowledge on the internal structure and hydrodynamical state.  In
turn, simulations are helping us understand how the intracluster medium
depends on non-gravitational processes such as galaxy formation \cite{Nagai07}.
The observations have made it clear that spectacular shocks permeate the intergalactic
medium. Simulations are able to connect this to the past history of
the gas \cite{Markevitch07,Skillman08}.  Structure, such as the
presence of a cooling core, depends upon merger history \cite{Burns08}.

\begin{figure}[t!]
\includegraphics[width=\textwidth]{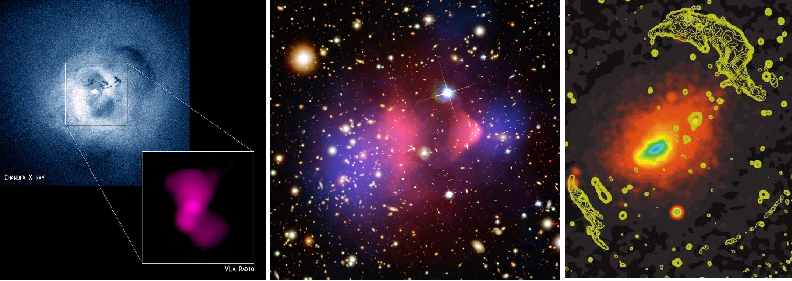}
  \caption{\label{fig:perseus} Cluster astrophysics at work! 
    {\em Left:} The {\sc Chandra} X-ray image of the Perseus 
    cluster showing bubble and shock substructure. 
    The VLA 90cm radio (inset) traces jets from the AGN
    filling the X-ray cavities \protect\cite{Fabian06}.
    {\em Center:} The ``bullet'' cluster 
    experiences a major merger that causes the gas (X-ray in red) to lag
    behind the Dark Matter (weak lensing mass in blue) \protect\cite{Clowe06}. 
    {\em Right:} Post-merging cluster A3667 (X-rays in color) with two giant radio
    synchrotron relics (contours) indicating the shock acceleration of 
    electrons \protect\cite{Rottgering97}.
    Parts of image composites courtesy {\sc Chandra} image archive.
  }
\end{figure}

\noindent\underline{\it Non-thermal processes:} 
In cluster cores, X-ray and
radio observations 
show clear relations between X-ray substructure and radio emission from
synchrotron-emitting particles injected by AGN \cite{Forman07,Sijacki08}.  See
Fig.~\ref{fig:perseus}. On large scales, some aspects of the diffuse
non-thermal emission in the radio and hard X-rays are still not understood
although considerable theoretical progress has been made by following cosmic
ray particle injection and transport while predicting the associated
non-thermal emission \cite{Pfrommer08}. 

\begin{figure}[t!]
  \includegraphics[width=6.5in]{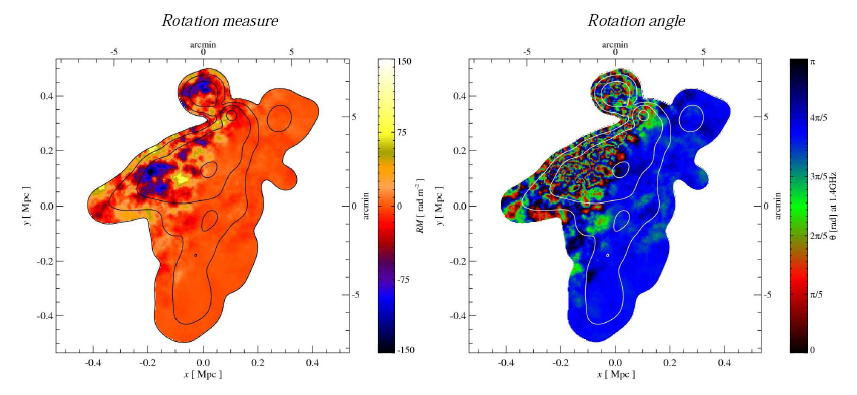}
  \caption{\label{fig:batt} 
    Simulated rotation measure (left) and 1.4GHz rotation
    angle (right) for an A2256-like cluster.  Detailed 
    study of magnetic field structure will be possible in the
    coming decade with new and improved low-frequency radio
    arrays.  Taken from Figure~5 of \protect\cite{Battaglia08}.
    }
\end{figure}

\noindent\underline{\it Magnetic Fields:} 
The new generation of detailed hydrodynamic
simulations are allowing us to start to model in detail the role of
magnetic fields and particles in clusters both in the bulk of the
intracluster medium as well as at interfaces of merging cores and
rising AGN bubbles.  This can be probed via sensitive low-frequency
radio observations of the synchrotron emission, and measurements of
the Faraday rotation \cite{Battaglia08}.  See Figure~\ref{fig:batt}.
The challenges include modeling associated anisotropic transport
processes of plasma particles that might govern the evolution of
magnetic fields through instabilities, the uncertain topology of
magnetic field structures, and the role of reconnection.

\noindent\underline{\it Sub-grid Physics:} 
There are a number of small-scale ``sub-grid'' 
astrophysical processes that impact the larger-scale structure of 
clusters.  In addition to the non-thermal physical effects noted
above, star formation also affects clusters at some level through
its feedback effect on galaxy formation.  Simulations are still in
the infant stages on modeling this, but progress is being made 
(e.g.\ \cite{Gnedin08}).

\noindent\underline{\it Golden Bullets:} 
The ``bullet'' cluster 1ES0657-558 is a 
spectacular example where detailed observations in the X-ray
and O/IR bands, plus weak gravitational lens modeling, are combined
to constrain fundamental physics such as the nature of gravity,
and Dark Matter cross-sections \cite{Clowe06,Randall08}.  


\section{New Horizons}

The preceding summary was meant to give a flavor of the high level of
activity in the field and the excitement that we feel looking ahead to
the future.  In the coming decade and beyond, we anticipate that even
greater progress can be made in key areas related to cluster
astrophysics.  These would be enabled by a number of initiatives in
various areas (with reference to companion white papers where they
are known to us):

\noindent\underline{\it Theory and Modeling Refinements:} Much of the
progress made in understanding cluster astrophysics has been based on
the tremendous improvements that have occurred in computational
resources and algorithm development. The next challenge will be to
advance our theoretical understanding of the micro-physics associated
with plasma non-equilibrium processes such as shocks, turbulence, and
magnetic dynamos and to connect those to astrophysical processes such
as accretion disks, jet and star formation.  Mapping the relevant
physics within cosmological simulations, for example to track the
baryonic (stellar and non-stellar) components \cite{WP-Kravtsov}, will
require improvements in dynamic range and simulation fidelity.
Understanding the results and limitations of simulations, carefully
simulating the next generation of observational data sets for
comparison, and the accurate extraction of astrophysical and
cosmological parameters from large simulation runs, will all be
challenges that must be met.  These goals can be achieved by
investment in the theory programs, and support for the development of
innovative new algorithms and implementation in the future code base.

\noindent\underline{\it Universe in a Box:} Realization of models in
the form of large and/or detailed simulations will build upon the
theory and modeling advancements.  Much of the future progress will
rely upon use of ``top-500'' class computational facilities.  It is
critical that astrophysical and cosmological researchers have funding
for the deployment of and access to the cutting edge computational and
access grid infrastructure of the future.  Results from simulations
will need to be accessible by the theorists and observers, in forms
that can be efficiently analyzed.  This is a trend across science
disciplines, and participation in cross-cutting programs with
computing centers and funding divisions will likely be necessary to
ensure that astrophysical computing needs are met.

\noindent\underline{\it Observational Frontiers:} 
A vigorous and broad program of multi-wavelength surveys and new
observational facilities will be needed to exploit the advances in
theory and modeling.  
Radio facilities at meter and centimeter wavelengths will discover and image
the synchrotron radiation from active and relic regions within
clusters \cite{WP-Rudnick}.  Substantial increases in sensitivity and
resolution at the lowest frequencies are particularly needed to go
deeper as well as to improve upon the first-generation radio sky
surveys.
Millimeter and submillimeter observations will probe the thermal pressure and
velocity structure through the SZ effect \cite{WP-Golwala},
with high-resolution observations pinpointing active star formation.
A combination of wide bandwidth large-format survey cameras and
high-resolution high-sensitivity interferometric arrays provide the
required capabilities.  Optical and infrared surveys of galaxies
will follow the stellar content, and through weak lensing map out the Dark
Matter to increasing precision.  The successors to the 
SDSS will be carried out using the next generation of ground and
space-based O/IR survey telescopes outfitted with large state-of-the-art
cameras, feeding follow-up programs using multi-object spectrographs
on large apertures.
X-ray observations will carry out calorimetry of the ICM, 
image shocked substructure \cite{WP-Vikhlinin} and diffuse cluster
emission \cite{WP-Markevitch}, and further our
understanding of AGN feedback \cite{WP-McNamara}. It is critical to
have future X-ray missions to follow {\sc Chandra} and {\sc XMM-Newton}, 
and to compile the next all-sky cluster survey.
Finally, high energy X-rays
and Gamma rays can be used to constrain Dark Matter through annihilation
emission, and to probe the high energy tail of shock and jet accelerated
particles.  
The combination of all of these in the acquisition of
extremely large samples of clusters for cosmology will enable
precision measurements of parameters such as the Dark Energy equation
of state complementary to those provided by fine-scale
CMB \cite{WP-Page} and optical surveys. 


\section{Conclusions}

It is clear that our understanding of cluster astrophysics has blossomed in the
past decade due to the convergence of theoretical, computational, and
observational advancements.  In turn, the recognition that cluster surveys can
be used for cosmology has kicked off a new generation of ambitious
observational programs on a variety of facilities.  The success of
these programs will
hinge on our ability to understand the nature and evolution of clusters.  We
have outlined key questions and focus areas in this subject for concentration
in the coming decade.  We support a strategy that leverages investment in
theory programs, progress in computational infrastructure and improved access
to computing and data products, and the adoption of a science road map that
makes use of precursor research from key near-term projects and pathfinder
telescopes, as well as engenders the design, construction and operation of the
important next-generation facilities (both small and large). 
The science program for cluster astrophysics and cosmology truly spans the
electromagnetic spectrum, crossing boundaries between modeling and
observing techniques, and is thus an ideal focus for the coming decade.




\bibliographystyle{unsrt}

\end{document}